\documentclass{aa}  

\usepackage{graphicx}
\usepackage{txfonts}
\usepackage{lipsum}
\usepackage{subcaption}         
\usepackage{lscape}             
\usepackage{placeins}           

\usepackage{booktabs}
\usepackage[usenames,dvipsnames]{xcolor}
\usepackage[normalem]{ulem}
\usepackage{enumitem}
\usepackage{url}
\usepackage[colorlinks=true,
    linkcolor=blue,
    filecolor=magenta,      
    urlcolor=blue,
    citecolor=blue]{hyperref}

\usepackage{xcolor}
\usepackage{ulem}

\usepackage{caption}
\newcommand*{\Msun}{\ensuremath{{\rm M}_{\odot}}}

\newcommand{\lsun}{\ifmmode L_{\odot} \else L$_{\odot}$\fi}
\newcommand*{\MESA}{\texttt{MESA}}
\newcommand*{\PyMMAMS}{\texttt{PyMMAMS}}
\newcommand*{\starm}{\texttt{star munchies}}

\def\msun{\ifmmode M_{\odot} \else M$_{\odot}$\fi}
\def\msunyr{\ifmmode M_{\odot} {\rm yr}^{-1} \else M$_{\odot}$ yr$^{-1}$\fi}
\def\zsun{\ifmmode Z_{\odot} \else Z$_{\odot}$\fi}
\newcommand{\mstar}{\ifmmode M_\star \else $M_\star$\fi}
\newcommand{\tgri}{\ifmmode t_{\rm GRI} \else $t_{\rm GRI}$\fi}
\newcommand{\mgri}{\ifmmode M_{\rm GRI} \else $M_{\rm GRI}$\fi}
\newcommand{\lgri}{\ifmmode L_{\rm GRI} \else $L_{\rm GRI}$\fi}
\newcommand{\macc}{\ifmmode \dot{M}_{\rm acc}  \else $\dot{M}_{\rm acc}$\fi}

\begin{document}


   \title{Ballistic mixing: stellar mergers drive internal mixing in extremely massive stars, with implications for proto-globular cluster pollution}

   \titlerunning{Stellar mergers drive internal mixing in extremely massive stars, with implications for proto-globular cluster pollution}

%
%
%

   \author{J. Roman-Garza\inst{1,2}\corrauth{jaime.romangarza@unige.ch}        
   \and C. Charbonnel\inst{1,2,3}
   \and A. Rantala\inst{4,5}
   \and T. Fragos\inst{1,2}
   \and M. Gieles\inst{6,7,8}
   \and L. Ramírez-Galeano\inst{1}
        }

    \authorrunning{Roman-Garza et al.}

    \institute{Département d’Astronomie, Université de Genève, Chemin Pegasi 51, CH-1290 Versoix, Switzerland 
                \and Gravitational Wave Science Center (GWSC), Université de Genève, CH-1211 Geneva, Switzerland.
                \and CNRS, IRAP, 14 Avenue E. Belin, 31400 Toulouse, France
                \and Institute of Astronomy, University of Cambridge, Madingley Road, Cambridge CB3 0HA, UK
                \and Kavli Institute for Cosmology, Cambridge (KICC), University of Cambridge, Madingley Road, Cambridge CB3 0HA, UK
                \and ICREA, Pg. Lluís Companys 23, E-08010 Barcelona, Spain
                \and Institut de Ciències del Cosmos (ICCUB), Universitat de Barcelona (UB), c. Martí i Franquès, 1, E-08028 Barcelona, Spain
                \and Institut d’Estudis Espacials de Catalunya (IEEC), Edifici RDIT, Campus UPC, E-08860 Castelldefels (Barcelona), Spain
                }

   \date{Received XX, 20XX}

 
  \abstract{
  Extremely massive stars, with masses of $10^3$--$10^4$~\Msun, have been proposed as potential sources of the chemical enrichment associated with multiple populations in globular clusters. 
  Early on the main sequence, these stars reach central temperatures of 
  65 to 75~MK, enabling nucleosynthesis through the CNO, NeNa, and MgAl cycles,  producing abundance patterns similar to those observed in the so-called cluster's second population. Stellar mergers are among the leading formation channels for extremely massive stars. 
  In this work, we follow the formation of a $\sim6700$~\Msun\ star via 253 stellar mergers, drawn from a direct $N$-body star cluster simulation with properties similar to the densest star clusters observed at redshift $z\sim10$. We evolve each star as a 1D stellar model and construct the merger products from mixing and shock heating prescriptions calibrated to 3D simulations of parabolic collisions. We find that entropy-based mixing efficiently transports nuclear-processed material from the convective core to the envelope, while merger-driven heating promotes chemical homogenization. As a result, the stellar surface becomes enriched with H-burning products, producing abundance patterns consistent with those observed in globular cluster multiple populations and supporting the role of merger-formed extremely massive stars as potential sources of early chemical enrichment in proto-globular cluster-like conditions.
  }


   \maketitle
   \raggedbottom
   \nolinenumbers

\section{Introduction}

The upper end of the massive-star mass spectrum may host extremely massive stars (EMSs; $10^3$--$10^4$~\Msun) and super-massive stars (SMSs; $\geq 10^4$~\Msun) \citep[following][]{2025MNRAS.544..483G}. Though hypothetical, these objects are of interest for their potential role in producing the Na--Al correlation and C--N, Na--O, and Al--Mg anti-correlations observed in multiple stellar populations (MSPs) of globular clusters (GCs) \citep{2006A&A...450..523C,2009A&A...505..117C,2009A&A...505..139C,2018ARA&A..56...83B}. These patterns trace the products of the CNO, NeNa, and MgAl cycles  requiring hydrogen burning at non-zero metallicity and central temperatures of $\sim 65$--$75$~MK \citep{2017A&A...608A..28P,2025MNRAS.544..483G}. Multiple-population GCs also show very limited helium spreads of at most $\Delta Y \sim 0.2$ \citep[in mass fraction;][]{2018ARA&A..56...83B,2005ApJ...631..868D}, pointing to a young stellar polluter \citep{2014MNRAS.437L..21D}. 
\citet{2017A&A...608A..28P} show that EMS early main sequence (MS) nucleosynthesis can reproduce the Na/Al enhancements and O/Mg depletion in NGC 2808 while keeping $\Delta Y \leq 0.15$. SMSs, proposed as primary GC polluters \citep{2014MNRAS.437L..21D,2018MNRAS.478.2461G}, instead reach central temperatures $\gtrsim 80$~MK and can underproduce Na and Mg while overproducing Al relative to observations \citep{2025MNRAS.544..483G,2025A&A...699A.223R}. Previous non-rotating EMS and SMS models attain full convection only via non-physical enhancement of convection efficiency; however under standard mixing length theory these stars are not fully convective, leaving open the question of how processed material reaches the stellar surface to pollute their environment while they are still on the early MS \citep{2014MNRAS.437L..21D,2025A&A...699A.223R}. Rotation can produce mixing to promote chemical homogenization in stars \citep[e.g.,][]{2000ARA&A..38..143M}, but the rotation rate of stars depends on their formation scenario \citep[][]{2012ApJ...748...97R} and it remains uncertain for collisionally assembled EMSs.

Stellar mergers are expected to play a key role in the formation and evolution of EMSs in dense stellar environments (i.e. $\gg10^6$~\Msun\,pc$^{-3}$), where repeated collisions can produce stars far more massive than their progenitors \citep[][]{2002ApJ...576..899P,2004Natur.428..724P,2018MNRAS.478.2461G,2025MNRAS.544..483G,2024Sci...384.1488F,2024MNRAS.531.3770R,2026MNRAS.549ag986R,2025A&A...704A.321V,2025MNRAS.543.1023L}. Hydrodynamic simulations show that merger remnants differ from stars formed through standard evolution: orbital energy dissipation partially inflates their structure, while induced mixing creates non-standard chemical profiles \citep[][]{2007ApJ...668..435S,2008MNRAS.383L...5G,2023MNRAS.519.5191B,2026A&A...707A.163R}. However, although stellar mergers are predicted to be common in dense stellar systems \citep[e.g][]{2024MNRAS.531.3770R,2026MNRAS.549ag986R}, their remnants are rarely evolved self-consistently. Instead, they are typically replaced by standard stellar models.

In this work, we present the evolution of an 1D stellar model built through successive mergers according to the collision histories from the dynamical simulations by \cite{2026MNRAS.549ag986R}. We focus on the role of mergers in the mixing of the resulting EMS and the yields of the mass that can be lost trough collisional mass loss and winds.

\section{Methods}
To study the formation of EMSs via stellar collisions, we developed the module "\starm", which couples the stellar evolution software instrument “Modules for Experiments in Stellar Astrophysics” \citep[\MESA;][]{mesa1,mesa2,mesa3,mesa4,mesa5,mesa6} with the python version of "Make Me a Massive Star" module \citep[\PyMMAMS;][]{2025arXiv251214669H,2008MNRAS.383L...5G}. The \starm\ module creates a pair of \MESA\ models with the masses of the primary and secondary stars that merge, $M_1$ and $M_2$ respectively, and calculates their evolution from the pre-MS up to the time when the collision occurs ($t_{\rm merger}$) according to the merger history. The internal profiles of mass, density, temperature and isotope abundances, of both primary and secondary stars just before the collision are then passed to \PyMMAMS\ to construct the structure of the corresponding merger product. The merger product is assembled following the entropy-sorting mixing prescription by \cite{2008MNRAS.383L...5G} calibrated by 3D simulations of head-on stellar collisions with zero initial orbital energy (same as a initial parabolic orbits with zero impact parameter). This prescription agrees with past and recent 3D stellar merger simulations \citep{2007ApJ...668..435S,2023MNRAS.519.5191B}. Additionally, we employ the shock-heating prescription by \cite{2025arXiv251214669H} to describe the energy introduced by the collision on the merger product. The resulting stratification in both entropy and chemicals within the merger product, induced by the heating and mixing prescriptions,
triggers the post-coalescence readjustment of its  structure, potentially driving further mixing by convection and/or thermohaline mixing. We do not consider the effect of rotation in our models to focus on the role of merger-induced mixing.

We follow the merger history of a single collisionally grown EMS in the simulations of \cite{2026MNRAS.549ag986R} for a cluster with central density $\sim5\times10^7 M_\odot$ pc$^{-3}$, $\Sigma_{\rm h}\sim 10^6$~\Msun\, pc$^{-2}$ and metallicity of $Z=0.01$~\zsun\ (model HD9), representative of the densest star clusters from the Cosmic Gems sample, observed at redshift $z\sim10$ \citep[][]{2024Natur.632..513A}. The simulation, performed using the \texttt{BIFROST} \textit{N}-body code \citep{2023MNRAS.522.5180R} models the rapid (<10 Myr) assembly of a dense million solar mass star cluster. 
For stellar evolution \texttt{BIFROST} is coupled to the fast population synthesis code \texttt{SEVN} \citep{2023MNRAS.524..426I}. All stellar collisions, most occurring on nearly parabolic orbits, result in mergers. The \textit{N}-body model also includes mild rejuvenation, mass loss in collisions \citep{2013MNRAS.434.3497G} and via EMS winds \citep[][]{2018A&A...615A.119V}. Our stellar evolution simulations do not include mass loss to avoid the uncertainties of these prescriptions, this results in a slight inconsistency in the time between collisions and the mass growth rate withing the $N$-body simulation and ours. However,  this does not affect  our conclusions regarding collision-induced mixing,  which is the focus of this work. Further details on the physical assumptions are provided in Appendix~\ref{app:physics_assumptions}.

\vspace{-0.9\baselineskip}
\section{Results}

The merger tree includes 2040 collisions; we model only the first 253 as extending the calculation would be more time-consuming without changing our conclusions. By then the star reaches $\sim6700$~\Msun. As the merger remnant increases in mass, its core temperature rises and the resulting nucleosynthesis is consistent with \cite{2017A&A...608A..28P}. The first merger, between stars of 93.6 and 91.8~\Msun, takes place at $\sim0.103$~Myr, producing a 185.4~\Msun\ remnant with a central temperature of $T_{\rm c}\sim56$~MK. At this point the star has reached CNO equilibrium and the NeNa cycle becomes significant, leading to a decline in central $^{20}$Ne and an increase in $^{23}$Na. The MgAl cycle is fully established by $\sim5000$~\Msun\ with $T_{\rm c}\gtrsim67$~MK, after about $\sim0.35$~Myr, which lowers the central $^{24}$Mg abundance while boosting $^{27}$Al. By the end of the simulation, the primary reaches $6744$~\Msun\ with $T_{\rm c}=68.1$~MK, central mass fractions $X(^1{\rm H})=0.645$ and $X(^4{\rm He})= 0.355$, and a helium enrichment of $\Delta Y=0.114$. If mass loss is implemented as in the $N$-body simulation, by merger 253 the collisional loss is $\sim700$~\Msun\ and wind loss $\sim1500$~\Msun, leaving $\sim4500$~\Msun. The wind mass loss is estimated by extrapolating the fitting formula by \cite{2018A&A...615A.119V} beyond its 900~\Msun\ fitting-range limit and may be overestimated. For further details on the nucleosynthesis of the merger product and the masses involved in the mergers refer to Appendix~\ref{app:nucleo}.

 \begin{figure*}
   \centering
    \includegraphics[width=\textwidth]{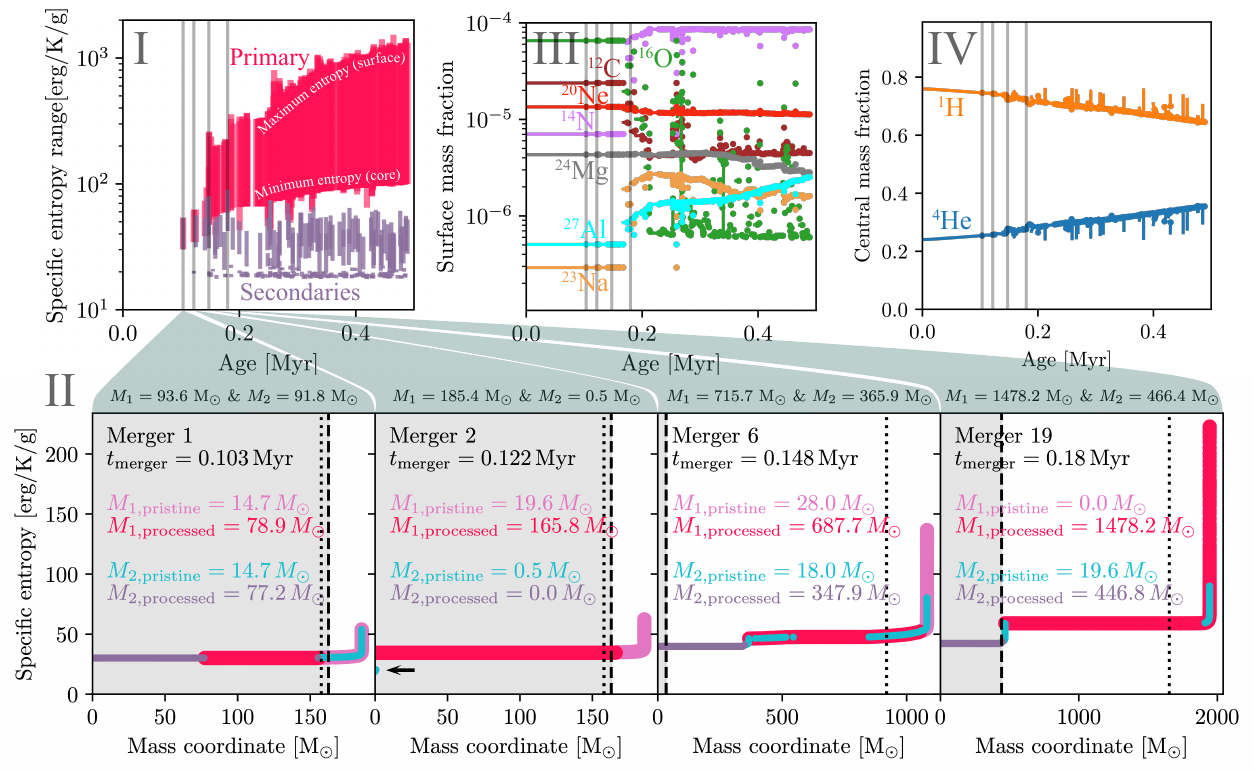}
   \caption{(Panel I) Time evolution of the specific entropy range, from minimum  (core) to maximum (surface), for the primary (magenta) and secondaries (purple) at the corresponding merger time.   
   (Panel II) Structure of the merger products for collisions 1, 2, 6 and 19, with the material of each shell from both stars color coded according to whether it retains the initial composition (pristine) or has been nuclear processed. The gray shaded regions show the convective core mass of the merger product just before the next collision, while the dotted line marks 85\% of its total mass. (Panel III) Time evolution of the surface abundances of key isotopes. 
   (Panel IV)  Evolution of the central hydrogen and helium mass fractions. Dots in Panels II and III represent the respective values at the time of the merger. The vertical solid lines in panels I, III and IV indicate the times of mergers 1, 2, 6 and 19. 
   }
   \vspace{-1.2\baselineskip}
   \label{fig:mixing_entropy}
\end{figure*}


Panel~I of Fig.~\ref{fig:mixing_entropy} illustrates the entropy range, from the core to the surface, 
of the primary and all the secondaries at the time of each collision along the merger tree. This comparison provides a first-order indication of where the secondary material will settle within the merger product given the prescription by \citet{2008MNRAS.383L...5G}: 
material with lower entropy sinks deeper into the primary star, while higher-entropy material is deposited at larger radii. A more detailed view of the mixing process is presented in panel~II, which shows the specific entropy profiles immediately after mergers 1, 2, 6, and 19. The figure distinguishes between pristine material (i.e., original composition when the stars were formed) and material already chemically processed by nuclear burning, i.e. the mass fraction of $^1$H is below 99\% of its initial value, in both the primary and secondary. The shaded area marks the convective core at the next collision, after the primary has thermally relaxed (fully or partially, depending on the delay between two consecutive collisions; see Appendix~\ref{app:merger_induced_conv}).

The first merger is representative of collisions with mass ratios close to unity, as $M_1 = 93.6$~\Msun\ and $M_2 = 91.8$~\Msun. Both stars have a the same age and similar entropy profiles. Consequently, the merger product inherits a core composed primarily of processed material from the cores of both progenitors. Only a small amount ($\sim5$~\Msun) of pristine material from the secondary is deposited within the convective core of the resulting merger (shown by the shaded region). In this case indeed, the merger product thermally relaxes before the next collision occurs, allowing it to reestablish convection in $\sim 85 \%$ of its mass.

 The second merger illustrates the opposite extreme in terms of mass ratio, with $M_1 =85.4$~\Msun\ and $M_2=0.5$~\Msun. The entropy throughout the secondary is lower than even the entropy of the primary's core, causing the entire secondary to sink to the very center of the merger product. Since this pre-MS low-mass star has undergone no nuclear processing, it contributes hydrogen-rich material to the core. The added pristine material amounts to only $0.3\%$ of the primary’s mass, such collisions with low-mass stars contribute to rejuvenating the merger product. The entropy profile also shows a layer of nuclearly processed material by the primary and which is located just above the convective core of the post-merger primary. The temperature in this radiative region is about $\sim22$~MK, which is sufficient for the H-burning through the CNO-cycle outside the fully mixed core.

As consecutive mergers produce mixing patterns analogous to those of the first two cases, we instead focus on the sixth merger, with $M_1 = 715.7$~\Msun\ and $M_2 = 365.9$~\Msun. Here, the core and a small fraction of the secondary's envelope have lower entropy than the primary's core. During the entropy-sorting procedure, this material is deposited at the center of the merger product, displacing nuclear-processed material from the primary outward in mass coordinate. In mergers 1 and 2, the convective core of the primary could retrieve $\sim 85 \%$ of the resulting merger's mass before the occurrence of the next collision (panel~III, Fig.~\ref{fig:mixing_entropy}). In contrast, the convective core retrieved after 
merger 6 has a smaller mass  due the interval between mergers 6 and 7 ($\lesssim 10^3$~yr) being shorter than the thermal timescale of the remnant (Appendix~\ref{app:merger_induced_conv}).  
 Relative to the typical convective $85\%$ mass, approximately $120$~\Msun\ of nuclear-processed material from the primary has been displaced into the envelope before merger 7, although it has not yet reached the surface. Finally, the nineteenth merger, with $M_1 = 1478.2$~\Msun\ and $M_2=466.4$~\Msun, illustrates the case in which successive mergers have fully displaced processed material of the primary throughout its envelope, enriching its surface with H-burning products.

Across the multiple mergers, the primary's envelope is not permanently convective; the merger-driven heating temporarily establishes convection there for about one thermal timescale, leading to some dredge-up of inner material towards the surface. In addition, molecular weight inversions produced by the mergers when processed and unprocessed material are deposited in adjacent layers can drive thermohaline mixing, although this process is generally less significant than convection (Appendix~\ref{app:merger_induced_conv}). As the processed material is displaced from the core outwards by the mergers and is mixed throughout the envelope by this transient convection and thermohaline mixing, the surface abundances begin to reflect the H-burning nucleosynthesis patterns shortly after 0.17~Myr, as shown in panel~III of Fig.~\ref{fig:mixing_entropy}. In comparison, the surface abundances of a 6744~\Msun\ standard stellar model remain unchanged (Appendix~\ref{app:comparison_model}).

Throughout the merger sequence, most secondaries are low-mass pre-MS stars that have undergone no or very little H-burning and bring pristine material towards the primary's core, inducing its partial rejuvenation as depicted by the slight rises in the central hydrogen mass fraction after the corresponding mergers in panel~IV of Fig.~\ref{fig:mixing_entropy}. In this simulation, however, this effect is relatively modest, as it contributes about $\lesssim 1\%$ of the primary’s core mass in unprocessed material.

\vspace{-0.8\baselineskip}
\section{Discussion and Conclusions}

We investigated the impact of consecutive stellar mergers on the mixing and chemical evolution of extremely massive stars. Following the collision history from \citet{2026MNRAS.549ag986R}, we model the formation of an EMS in a dense star cluster by evolving each progenitor with \MESA\ and constructing the merger products using the entropy-based mixing and shock-heating prescriptions respectively from \citet{2008MNRAS.383L...5G} and \citet{2025arXiv251214669H}. We show that consecutive mergers provide an efficient  mechanism not only to assemble EMSs, but also to modify their internal structure and surface composition through what we propose to call "ballistic mixing". Entropy sorting during mergers redistributes material throughout the star, allowing nuclear-processed material from the core to reach the envelope, while merger-induced heating generates transient convection and thermohaline mixing that promote chemical homogenisation. Consequently, the EMS surface becomes enriched with the products of H-burning, including the CNO, NeNa, and MgAl cycles while undergoing very modest He enrichment.  
These abundance patterns are consistent with those observed in the multiple populations of GSs with this initial mass ($\sim10^6\,\msun$) and metallicity ($Z=0.01\,\zsun$) \citep[][]{2006A&A...450..523C,2009A&A...505..139C,2018ARA&A..56...83B}, as well with the cases that present Mg depletion as consequence of the MgAl cycle \citep[e.g.][]{2012MNRAS.426.2889M,2014A&A...564A..60C}. 
Additionally, collisions rejuvenate the merger product, contributing pristine material amounting to only about $\lesssim 1\%$ of the primary star’s core mass in the simulations presented here.

The predicted enrichment of N-rich material also provides a possible connection with the N-emitting sources observed at various redshifts \citep[e.g.][]{2023A&A...677A..88B,2026MNRAS.545f2110J,2026A&A...710A.321M} with young massive clusters, suggesting that merger-formed EMSs may contribute to early chemical enrichment in dense stellar environments at non-zero metallicities \citep{2023A&A...673L...7C,2024ApJ...966...92S}.

Our work expands on the simulations of \cite{2009A&A...497..255G}, which formed stars up to $\sim 1000$~\Msun\ trough consecutive stellar collisions,  but assumed each merger produces a chemically homogeneous remnant. This assumption was based on the biased interpretation of the results from the 3D simulations of \cite{2007ApJ...668..435S}. The latter study finds that for mergers with mass ratios near zero (typical for collisionally assembled EMSs) the mixing matches the prescription of \cite{2008MNRAS.383L...5G}, based on comparisons of initial and final composition and entropy profiles. Implementing this mixing prescription shows that ballistic mixing does not require full chemical homogenization as multiple mergers can yield a similar final structure. While consecutive mergers using the same prescription have been studied in the past \citep[e.g.][]{2008A&A...488.1017G}, the merger product’s final composition has not been analyzed in detail.

We note that the maximum mass reached by the EMS and the timescale over which it  growths in mass and undergoes the surface enrichment of the nucleosynthesis products will depend on the adopted merger tree, including the delay between consecutive collisions, and the mass of the secondaries. This assembly timescale is inversely proportional to the stellar collision cross section, which can be substantially increased by merger-induced expansion \citep[][]{2026A&A...707A.163R} or by gas accretion onto the EMS \citep[e.g.,][]{nandal2023critical}. It will also strongly depend on the actual mass and compactness of the host cluster, which are key parameters of the multiple stellar populations properties \citep{2009A&A...505..139C,2017A&A...601A.112P,2026MNRAS.549ag986R}. In Appendix~\ref{app:cluster_cond}, we show that reducing the cluster density by two orders of magnitude increases both the EMS assembly and merger-induced mixing timescales by one order of magnitude and yielding $\Delta Y=0.48$ that is $\sim2.5$ times the maximum observed value \citep[][]{2018ARA&A..56...83B,2005ApJ...631..868D}.

Lastly, it is expected that the EMS will lose mass, either through stellar winds or as a consequence of the collisions, which is not included in our stellar models to mainly focus on the merger-induced mixing and avoid the uncertainties related to such processes. For the metallicity considered here, \cite{2018A&A...615A.119V} estimates that stars with 900~\Msun\ experience a $\dot{M}=-10^{-4.5}$~\Msun/yr, while such value remains uncertain for more massive stars. In the case of collisions, \citet{2013MNRAS.434.3497G} estimate that $\lesssim10\%$ of the total mass is removed through each merger depending on the mass ratio and corresponds to $\sim700$~\Msun\ if the prescription is included in our simulation. Although other studies suggest that this fraction can be higher \citep[][]{2023MNRAS.519.5191B,2025A&A...699A.223R,2026A&A...707A.163R}. Mass loss would expose the nuclear burning products to the EMS surface earlier than predicted in this work, by pealing off the EMS outer layers. While these effects will not affect our conclusions on the ballistic mixing process, they should be included self-consistently in future stellar and star cluster simulations to establish reliable quantitative timelines for star cluster early pollution. Besides the role of mass loss, we expect to address in future work the role of rotation on the mixing of merger assembled EMSs.

\begin{acknowledgements}
      This work was supported by the Swiss National Science Foundation (project number CRSII5\_213497). MG acknowledges the grants PID2024-155720NB-I00 and CEX2024-001451-M funded by MCIN/AEI/10.13039/ 501100011033. This research was supported by the International Space Science Institute (ISSI) in Bern, through the Team project \#25-636 “The Origin of Multiple Populations in Globular Clusters”. AR acknowledges support from the Kavli Foundation. The authors thank the anonymous referee for the provided comments on this work.
\end{acknowledgements}

%
\vspace{-2.2\baselineskip}
\bibliographystyle{aa} 
\bibliography{references}







   
  



\begin{appendix}
\nolinenumbers




\section{Stellar evolution and merger history assumptions}
\label{app:physics_assumptions}

\begin{figure}
   \centering
    \includegraphics[width=3.0in]{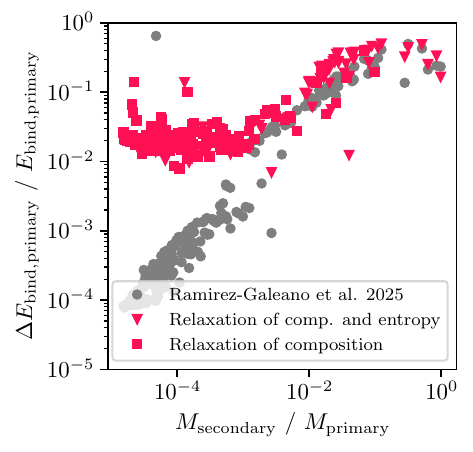}
   \caption{Ratio of the merger-injected energy ($\Delta E_{\rm bind,primary}$) to the binding energy of an undisturbed primary star ($E_{\rm bind,primary}$) as a function of the mass ratio. The magenta symbols indicate the values obtained from our simulations. Triangular markers correspond to models in which both the composition and entropy profiles from \PyMMAMS\ are used to construct the merger product in \MESA, whereas square markers indicate models where only the composition profile is employed. For comparison, gray circles show the expected heating from merger events based on the orbital energy loss prescription of \cite{2025A&A...699A.223R}.}
   \label{fig:delta_energy}
\end{figure}

To construct a \MESA\footnote{All stellar models presented in this study were computed using the \MESA\ software instrument (release 26.04.1) and the \texttt{MESA SDK} version 24.7.1.}\ model for the merger remnant, \starm\ module reloads the final model of the primary star into \MESA\ and adjust its mass $M_1$ to $M_{\rm merger}$ using the \texttt{relax\_mass} procedure over a timescale of 1~yr, i.e. $\lesssim1/10$ of the primary's dynamical timescale. \PyMMAMS\ assumes an ideal gas equation of state, whereas \MESA\ employs a blend between several tabulated EOS in different regimes of the $\log T$--$\log \rho$ plane \citep[see][for further context]{mesa1,mesa4,mesa6}. As well, \MESA\ accounts for physical processes that \PyMMAMS\ do not consider in order to predict the structure of the merger product. As a consequence of the previous differences, in the case where mass ratio of the merging stars is $\lesssim 0.01$ where merger induced heating is negligible, usually the \PyMMAMS\ merger product is more compact than the hydrostatic model obtained via mass relaxation of the primary in \MESA. To guaranty numerical stability, if the \PyMMAMS\ merger product has a radius smaller than the one of the mass-relaxed model, we treat shock heating as negligible and the code relaxes only the composition of the mass-relaxed model using the \PyMMAMS\ output. Otherwise, when the \PyMMAMS\ structure is not more compact, we relax both the composition and entropy profiles of the mass-relaxed model to match the \PyMMAMS\ result. Fig.~\ref{fig:delta_energy} shows that, on average, the cases where only composition is relaxed to create the merger product the resulting binding energy of such models is only $\sim 1\%$ higher with respect to the one of the un-perturbed model created trough the mass-relaxation only. For those cases where only composition is relaxed, as the time between consecutive mergers is $\Delta t > 10$~yr, that is bigger than $1\%$ of the primary's thermal timescale, the primary relaxes its structure before the next collision. After the relaxation steps, the resulting model is advanced forward in time until the next scheduled collision occurs, and the same procedure is then applied again to obtain the product of the subsequent merger.

In Fig.~\ref{fig:delta_energy}, we also compare our results with the heating prescription of \cite{2025A&A...699A.223R}, which is based on the orbital energy dissipated as the system evolves from an initially parabolic orbit to the point at which the secondary fills its Roche lobe. Overall, this prescription predicts less heating than the shock-heating prescription adopted in \PyMMAMS\ \citep{2025arXiv251214669H,2008MNRAS.383L...5G}.

All \MESA\ models are initialized with a metallicity of $Z=0.01$~Z$_\odot$ and a helium mass fraction of $Y=0.24$, similarly to the stellar tracks employed by \cite{2026MNRAS.549ag986R}. We do not include rotation in any of the models considered in this work. Angular momentum should be transfered from the secondary’s initial orbit to the primary during the merger and the resulting rotational profile might aid on the mixing of the merger product \citep[e.g.][]{2000ARA&A..38..143M}. But our framework cannot determine the resulting remnant’s rotation profile in a self-consistent manner as collisions are not confined to the equatorial plane as assumed by \MESA\ for the 1D stellar model approximation. Convection is treated using the mixing-length theory (MLT) of \cite{mihalas1978stellar}, adopting a mixing-length parameter of $\alpha_{\rm MLT} = 1.93$ \citep{2016ApJ...823..102C,2023ApJS..264...45F}. We note that in standard MLT implementations used in \MESA, the transport of chemical species is treated as a diffusive process rather than as instantaneous mixing, we provide a comparison with models using time dependent convection in Appendix~\ref{app:merger_induced_conv}. The emergence of convective regions is determined according to the Ledoux stability criterion. We adopt the value for the thermohaline mixing coefficient as $\alpha_{\rm th}=17.5$ \citep[corresponding for an aspect ratio of unity for the finger structures given the formalism by][]{2007A&A...467L..15C}. 

To track self consistently the nucleosynthesis of EMSs during the main sequence, we implement a costume nuclear network for our \MESA\ models \citep[the same used by ][]{2026arXiv260321714R}. The implemented nuclear network extends the \texttt{sagb\_NeNa\_MgAl} network from \MESA, it tracks 45 isotopes instead of 29, where now the most massive isotope followed consistently is $^{30}$P. The cumulative abundance of all non-tracked species is assigned to the abundance of $^{37}$Cl. The network follows the CNO, NeNa, and MgAl cycles, adds the reaction rates for the branching reactions from aluminum to silicon and phosphorus as well as all decaying reactions for $^{24}$Mg. In this study, we do not account for the impact of uncertainties in nuclear reaction rates on the resulting nucleosynthesis by the EMS.

The collisions experienced by the primary star in the $N$-body simulation have near unity eccentricity and different impact parameters. The merger induced mixing and heating prescriptions considered for this work are fitted to simulations of parabolic orbit with an impact parameter value of zero \citep[][]{2008MNRAS.383L...5G,2025arXiv251214669H}. Given the limitations of such prescriptions, we do not consider the effects of having collisions with different impact parameters.

\section{Details on the nucleosynthesis}
\label{app:nucleo}

\begin{figure}
   \centering
    \includegraphics[width=3.0in]{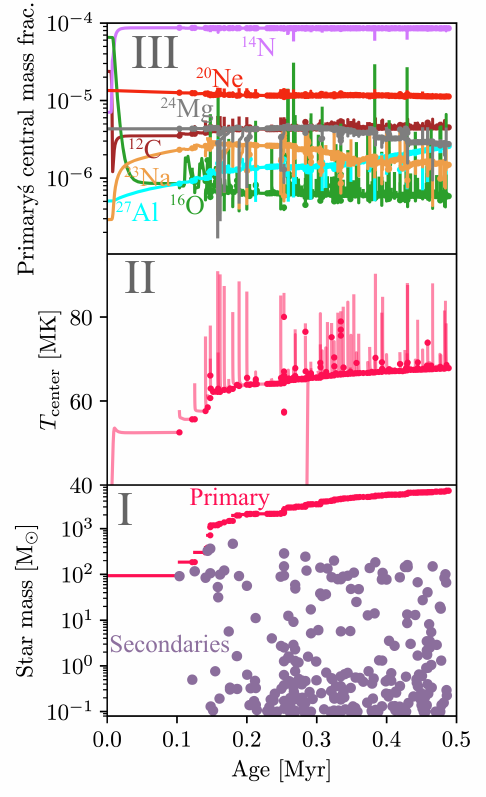} 
   \caption{Time evolution of: (I) the masses of the primary and secondary (magenta and purple, respectively); (II) the primary's central temperature, and (III) the primary's core mass fractions of $^{12}$C, $^{14}$N, $^{16}$O, $^{20}$Ne, $^{23}$Na, $^{24}$Mg and $^{27}$Al color coded. Dots in Panels II and III represent the respective values at the time of the merger.}
   \vspace{-1.2\baselineskip}
   \label{fig:nucleosynthesis}
\end{figure}

Panel I of Fig.~\ref{fig:nucleosynthesis} shows the evolution of the primary mass and the masses of the accreted secondaries. Panel II shows the evolution of the primary's central temperature as a function of time, and Panel III the evolution of its central abundances of $^{12}$C, $^{14}$N, $^{16}$O, $^{20}$Ne, $^{23}$Na, $^{24}$Mg, and $^{27}$Al, tracing the CNO, NeNa, and MgAl cycles (all the other relevant isotopes are included in the calculations but not shown in the plot).

Panel II of Fig.~\ref{fig:nucleosynthesis} shows the increment or decrement of the temperature of the innermost cell of our stellar models just after a merger happens. This is obtained as a product of the relaxation step to build the structure of the merger product, as the mass, composition and entropy are changed in much less than the dynamical timescale of the star. The increments and decrements of central temperature last $< 100$~yr, that is $<1/10$ of the primary's thermal timescale. Any drastic changes in the composition of the innermost cell as a product of the variation of the central temperature are quickly diluted as the merger product relaxes thermally and convection mixes the inner region of the star.

\section{Merger-induced convection and thermohaline mixing}
\label{app:merger_induced_conv}

 \begin{figure*}[h!]
   \centering
    \includegraphics[width=\textwidth]{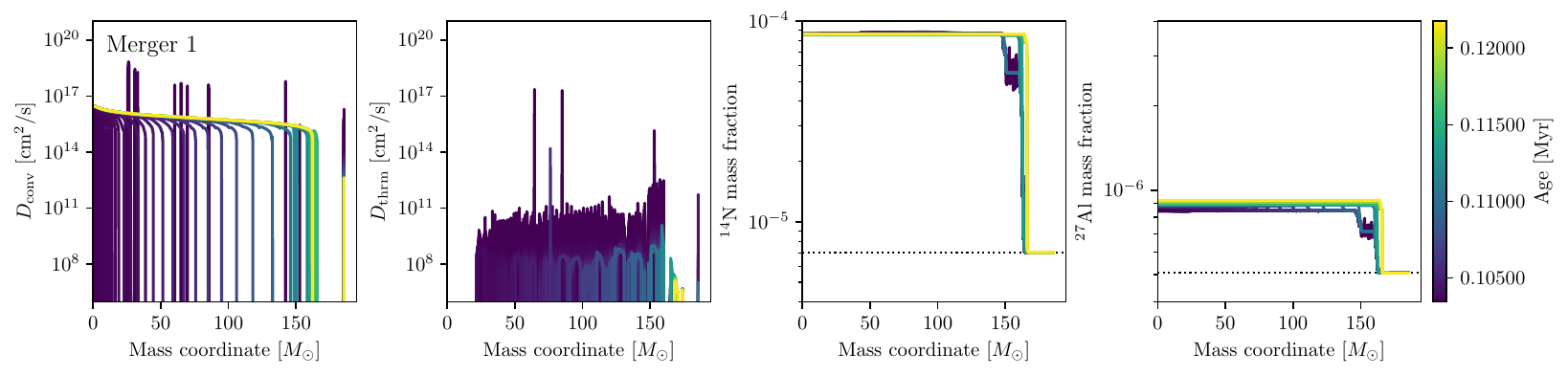}
    \includegraphics[width=\textwidth]{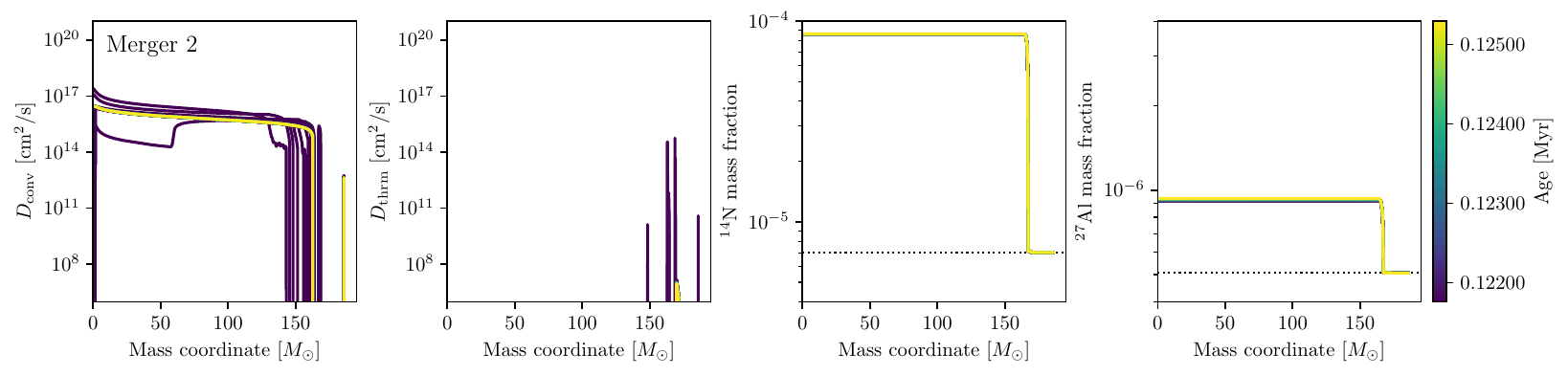}
    \includegraphics[width=\textwidth]{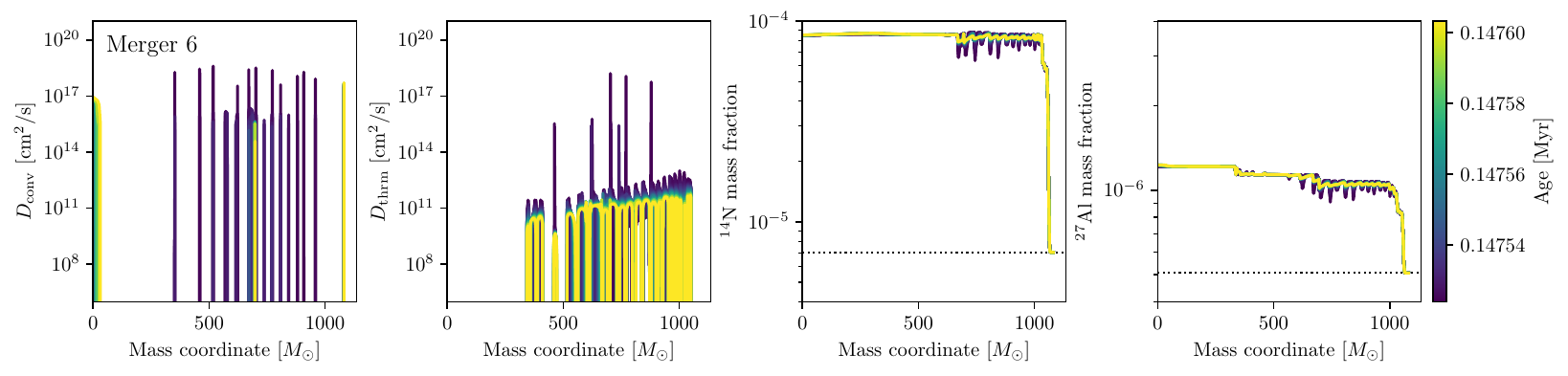}
    \includegraphics[width=\textwidth]{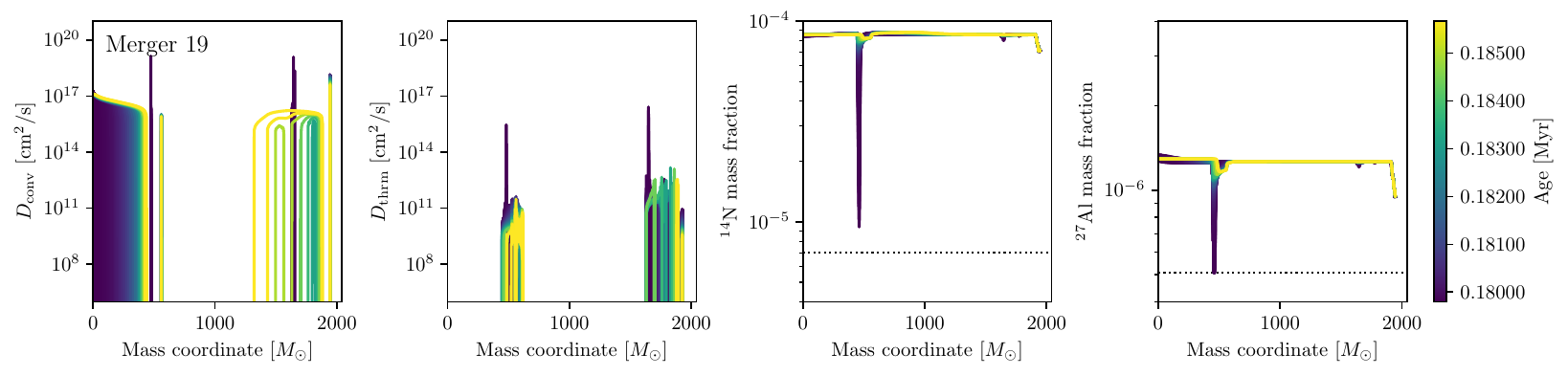}
   \caption{Time evolution of the post-merger profiles of the convective and thermohaline mixing diffusion coefficients, together with the mass fractions of $^{14}$N and $^{27}$Al, is shown as a function of the mass coordinate. Results are displayed for the remnants of mergers 1, 2, 6, and 19, in that order for the rows from top to bottom, and the evolution is tracked up to the onset of the next merger. The dotted lines in the $^{14}$N and $^{27}$Al mass-fraction panels indicate the pristine abundances of these isotopes.}
   \label{fig:conv_and_thrm}
\end{figure*}

Fig.~\ref{fig:conv_and_thrm} presents the time evolution of the stellar structure of each merger product, from the corresponding collision event up to the subsequent merger, for collisions 1, 2, 6, and 19 shown in separate rows. We display the convective and thermohaline mixing  diffusion coefficients, $D_{\rm conv}$ and $D_{\rm thrm}$, respectively, together with the mass fractions of $^{14}$N and $^{27}$Al as functions of the mass coordinate. Colors code the time since the moment of the merger. The first merger event produces a star of 185.4~\Msun, as $M_1 = 93.6$~\Msun\ and $M_2 = 91.8$~\Msun, there the energy deposited by shock heating corresponds to $\sim 17\%$ of the binding energy of an unperturbed star with the same mass. Consequently, convection in the core is shut off, and multiple convective zones appear throughout the central region. After a few $1000$~yr, comparable to the thermal timescale of this star, core convection is reestablished. Due to the chemical stratification of the merger remnant, together with the entropy and temperature gradients induced by shock heating, thermohaline mixing develops across both the core and the envelope; however, as the merger product reaches thermal equilibrium, the associated mixing coefficient becomes negligible. Regarding the $^{14}$N and $^{27}$Al abundances, we see that mixing progressively smooths the initial composition profile of the merger product, while at this stage the envelope still consists predominantly of pristine material.

For the second merger, in which the 185.4~\Msun\ star accretes a 0.5~\Msun\ companion, the \MESA\ model of the post-merger object is generated by relaxing only the composition profile according to the \PyMMAMS\ output, without adjusting the entropy (as described in Appendix~\ref{app:physics_assumptions}). Under this treatment, the relaxation in \MESA\ produces a star whose binding energy is higher by only $\sim 0.7\%$ from that of an unperturbed star with the same total mass. The second row of Fig.~\ref{fig:conv_and_thrm} demonstrates that, following the second merger, the diffusion coefficients and the abundance profiles show a negligible change throughout the subsequent evolution before the next merger happens.

The sixth merger, with $M_1 = 715.7$~\Msun\ and $M_2=365.9$~\Msun, yields a star with a mass of 1081.6~\Msun. Under the shock-heating prescription, this event deposits roughly 49\% of the binding energy of an unperturbed star of the same mass and evolution phase. The subsequent evolution of the mixing coefficients resembles that of the first merger; however, in this case the next collision occurs about $\sim 800$ yr later. During this interval, the star does not have sufficient time to thermally relax, so the convective core immediately before the next merger remains smaller than the typical $\sim 85\%$ of the stellar mass found in the other models. Up to this point in the evolution, the repeated mergers mix nuclear-processed material into regions outside the convective core, but the stellar surface has not been enriched yet.

The nineteenth merger, with $M_1 = 1478.2$~\Msun\ and $M_2=466.4$~\Msun, $M$ produces a star with a mass of $1944.6$~\Msun, where shock heating contributes roughly $44\%$ of the binding energy of an equivalent, unperturbed star. The subsequent evolution of the mixing coefficients resembles that found in merger 6, although in this case convection is reestablished in a larger portion of the core as the time to the next collision is $\sim6000$~yr. We find that convection zones develop throughout the star’s outer layers, which smooth out the initial composition profiles. At this stage, the processed material has already reached the stellar surface.

As convective and thermohaline mixing become effective only after the merger product begins to thermally relax, the primary mechanism responsible for transporting nuclear-processed material from the EMS core to the outer layers is the entropy-based mixing that occurs during the merger itself. Convection and thermohaline mixing mainly act to smooth the abundance gradients established by this process rather than producing the enrichment directly. 

To assess the impact of the treatment of convection, we performed an additional simulation identical to our fiducial model but employing the time-dependent convection (TDC) formalism based on the one equation model by \cite{1986A&A...160..116K}. The TDC formalism tracks the time evolution of the effective kinetic energy of turbulence and aims to present a framework for the transport of energy and chemicals in a more self-consistent manner with respect to their time dependence \citep[][]{mesa6}. Unlike the standard MLT implementation, where convective regions are assumed to adjust instantaneously to local stability criteria, TDC follows the temporal evolution of the formation and dissipation of convective zones. Analogously to the mixing-length theory (MLT) prescription adopted in our fiducial model, the time-dependent convection (TDC) formalism also treats the transport of chemical species as a diffusive process.

\begin{figure}
   \centering
    \includegraphics[width=3.in]{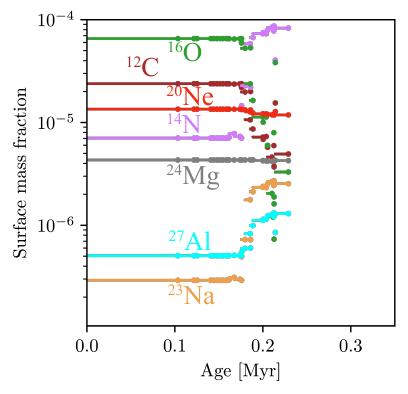}
   \caption{Variation of the surface abundances of $^{12}$C, $^{14}$N, $^{16}$O, $^{20}$Ne, $^{23}$Na, $^{24}$Mg and $^{27}$Al across time, for the simulation with TDC. }
   \label{fig:surf_abundance_tdc}
\end{figure}

Figure~\ref{fig:surf_abundance_tdc} shows the evolution of the surface mass fractions of the same representative nuclear products presented in Fig.~\ref{fig:mixing_entropy} for the TDC simulation. Due to convergence issues, the TDC calculation could only be evolved through the first 34 consecutive mergers, and therefore we do not adopt this prescription for our reference models. Nevertheless, despite the smaller number of completed mergers, the resulting surface chemical enrichment is consistent with that obtained using the standard MLT treatment. As the delay time for the primary's surface to show a similar enrichment as the one in the MLT simulation is $<10^{-3}$~Myr. Such indicates that our conclusions are not sensitive to the adopted convection prescription.

\section{Comparison with a non-merger product EMS}
\label{app:comparison_model}

\begin{figure} [h!]
   \centering
    \includegraphics[width=3.in]{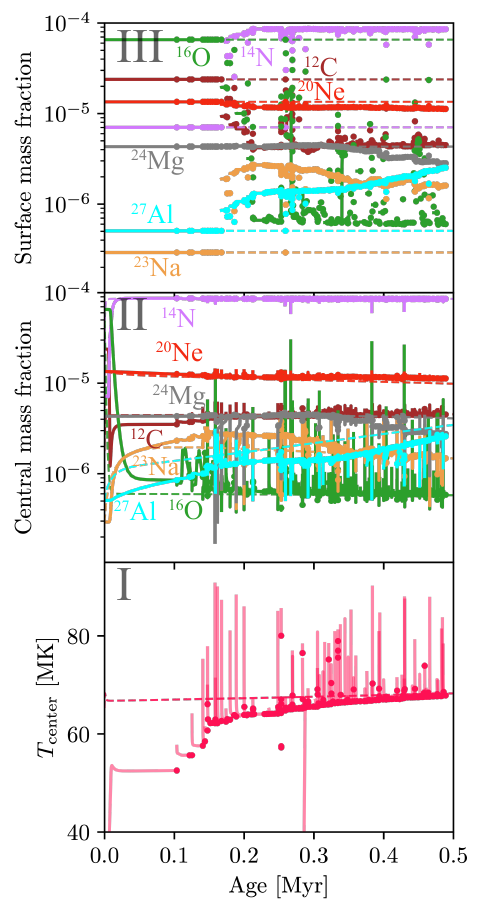}
   \caption{ (Panel I) Temporal evolution of the central temperature for the merger remnant in our fiducial model (solid lines) compared with that of a non-merger star with a mass of $6744$~\Msun (dashed line).  
(Panel II) Time evolution of the central mass fractions of $^{12}$C, $^{14}$N, $^{16}$O, $^{20}$Ne, $^{23}$Na, $^{24}$Mg, and $^{27}$Al for the merger remnant (solid lines) and the non-merger star (dashed lines).  
(Panel III) Time evolution of the surface mass fractions of the previous isotops for the merger remnant (solid lines) and the non-merger star (dashed lines).}
   \label{fig:comp_non_merger_product}
\end{figure}

In Fig.~\ref{fig:comp_non_merger_product} we present a comparison of the nucleosynthesis products and surface chemical enrichment between the merger product of our fiducial model and a 6744~\Msun\ star that does not undergo any collisions or mass loss, followed from the pre-MS phase to an age of 0.5~Myr. Panel I shows that the central temperature of the non-merger model remains approximately constant at $\sim 67$~MK. Consequently, the nucleosynthesis in this model resembles that of the merger product at later evolutionary stages, as illustrated in panel II, since at these temperatures the CNO, NeNa, and MgAl cycles are fully activated. Finally, panel III demonstrates that, unlike the merger product in our fiducial model, the surface of the non-merger EMS does not become enriched in H-burning products.

\section{Dependence on cluster conditions}
\label{app:cluster_cond}

To investigate the sensitivity of the EMS assembly and internal mixing to the properties of the host cluster, we considered an alternative merger history from \cite{2024MNRAS.531.3770R}, characterized by an initial central density approximately two orders of magnitude lower ($\rho_\mathrm{central}\sim5\times10^5 M_\odot$ pc$^{-3}$ vs $5\times10^7 M_\odot$ pc$^{-3}$) than that of the HD9 cluster from \citet[][]{2026MNRAS.549ag986R} adopted in our main analysis; as well the density is about one order of magnitude below the estimated one for the less dense star clusters in the Cosmic Gems sample by \cite{2024Natur.632..513A}. The initial metallicity is identical in both simulations. The cluster with higher initial density hosts a collision cascade that both begins earlier (first collisions at $0.1$ Myr vs $1.0$ Myr) and sustains overall higher stellar collision rates. We note that \cite{2024MNRAS.531.3770R} included simplified stellar merger and EMS wind treatments resulting in less collision and wind mass loss compared our fiducial set of stellar collisions.

\begin{figure}
   \centering
    \includegraphics[width=3in]{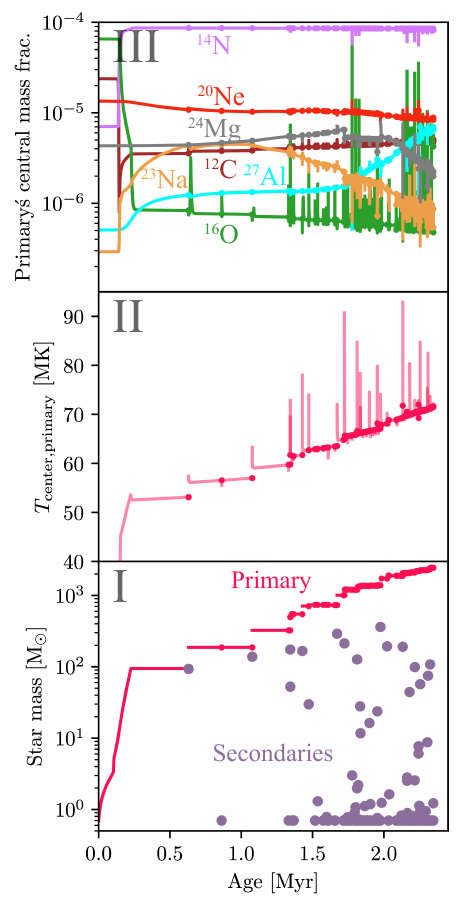} 
   \caption{Same as Fig.~\ref{fig:nucleosynthesis}, but corresponding to the simulation of the cluster with reduced density.}
   \label{fig:nucleosynthesis_other_tree}
\end{figure}

\begin{figure}
   \centering
    \includegraphics[width=3in]{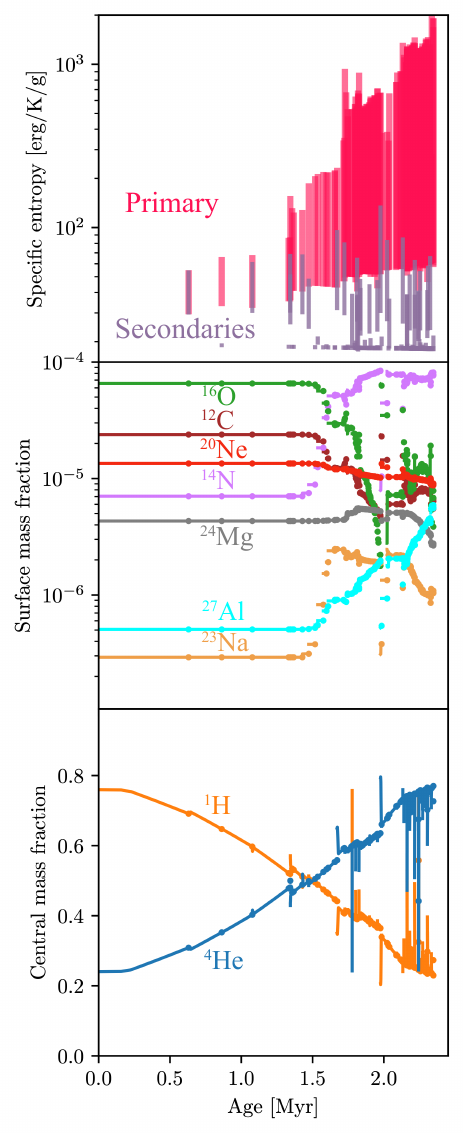}
   \caption{Identical to panels I, III, and IV of Fig.~\ref{fig:mixing_entropy} (from top to bottom, respectively), but corresponding to the lower-density cluster simulation.}
   \label{fig:mixing_entropy_other_tree}
\end{figure}

Figure~\ref{fig:nucleosynthesis_other_tree} is analogous to Fig.~\ref{fig:nucleosynthesis}, but refers to the lower-density cluster. The corresponding merger tree contains 177 collisions, all of which were successfully modeled with \starm\ without any convergence issues. The final merger remnant attains a mass of $2435$~\Msun\ by $\sim 2.3$~Myr, and is more advanced on the MS than in the simulation presented in the main text. Consequently, the central temperature could increase to $\sim 71$~MK due to higher molecular weight in the more He-rich core (Fig.~\ref{fig:mixing_entropy_other_tree}), allowing the MgAl cycle to become fully active \citep{2017A&A...608A..28P}, and leading to a depletion of $^{24}$Mg (with a concurrent decrease in $^{23}$Na) and a monotonic increase of the central mass fraction of $^{27}$Al by approximately one order of magnitude by the end of the simulation.

The slower assembly of the EMS also affects merger-induced mixing. Fig.~\ref{fig:mixing_entropy_other_tree} presents the analog of panels I, III, and IV from Fig.~\ref{fig:mixing_entropy}, but for the lower-density cluster simulation. In this configuration, the stellar surface does not become enriched in H-burning products until $\sim 1.5$ Myr, a timescale that is nearly an order of magnitude longer than in the HD9 cluster simulation. Moreover, the merger product in the lower-density cluster simulation is more evolved: by the end of the simulation it reaches central mass fractions of $^1$H and $^4$He of $\sim 0.27$ and $\sim 0.77$, respectively, corresponding to a helium enhancement of $\Delta Y = 0.48$. Such final helium enhancement is higher by a factor of $2.4$ with respect to the maximum observed values of $\Delta Y$ in GCs \citep[][]{2018ARA&A..56...83B,2005ApJ...631..868D}.

\end{appendix}
\end{document}